%Paper: gr-qc/9402001
%From: LBSZAB@rmk530.rmki.kfki.hu
%Date: Tue, 1 Feb 1994 16:09 GMT+1

%%%%%%%%%%%%%%%%%%%%%%%%%%%%%%%%%%%%%%%%%%%%%%%%%%%%%%%%%%%%%
%%%%%%%%%%%%%%%%%%%%%  PLAIN TEX FILE  %%%%%%%%%%%%%%%%%%%%%%
%%%%%%%%%%%%%%%%%%%%%%%%%%%%%%%%%%%%%%%%%%%%%%%%%%%%%%%%%%%%%

\font\llbf=cmbx10 scaled\magstep2
\font\lbf=cmbx10 scaled\magstep1

\def\edth{\hskip 3pt {}^\prime \kern -6pt \partial }

\hsize=15truecm
\vsize=22truecm
\parindent=1truecm
\baselineskip 14pt plus 2pt

\centerline{\llbf Two dimensional Sen connections}\par
\centerline{\llbf in general relativity}\par
\bigskip
\centerline{\bf L. B. Szabados}\par
\centerline{Research Institute for Particle and Nuclear Physics}\par
\centerline{H--1525 Budapest 114, P.O.Box 49, Hungary}\par
\centerline{E-mail: lbszab@rmki.kfki.hu}\par
\bigskip
\bigskip
\centerline{\bf Abstract}\par

\noindent The two dimensional version of the Sen connection for spinors
and tensors on spacelike 2-surfaces is constructed. A complex metric
$\gamma_{AB}$ on the spin spaces is found which characterizes both the
algebraic and extrinsic geometrical properties of the 2-surface $\$ $.
The curvature of the two dimensional Sen operator $\Delta_e$ is the pull
back to $\$ $ of the anti-self-dual part of the spacetime curvature
while its `torsion' is a boost gauge invariant expression of the extrinsic
curvatures of $\$ $. The difference of the 2 dimensional Sen and the induced
spin connections is the anti-self-dual part of the `torsion'. The
irreducible parts of $\Delta_e$ are shown to be the familiar 2-surface
twistor and the Weyl--Sen--Witten operators. Two Sen--Witten type identities
are derived, the first is an identity between the 2 dimensional twistor
and the Weyl--Sen--Witten operators and the integrand of Penrose's charge
integral, while the second contains the `torsion' as well. For spinor fields
satisfying the 2-surface twistor equation the first reduces to Tod's formula
for the kinematical twistor.\par
\vskip 1.5truecm

{\lbf Introduction}\par \vskip 0.5truecm

\noindent It is well known that one cannot associate gauge-independent
energy-momentum (and angular momentum) {\it density} with the gravitational
`field'; i.e. any such expression is pseudotensorial or, in the tetrad
formalism of gravity, depends on the tetrad field too. For asymptotically
flat spacetimes, however, one can define the {\it total} energy-momentum
[1-3]. One of the most important results of
the last decade in the classical relativity theory is the better
understanding of the energy-momentum of localized gravitating systems,
especially the proof of the positivity of the ADM and Bondi--Sachs
masses [4-8]. These results can naturally be recovered from the spinorial
Sparling equation if on the hypersurfaces extending either to spatial or
null infinity the 3 dimensional Sen connection is used [9,10]. This Sen
connection seems therefore to be the `natural' connection on these
hypersurfaces in the energy-momentum problems of gravity. \par
    These successes inspired several relativists to search for expressions
of the gravitational energy-momentum at the {\it quasi-local} level; i.e.
to associate these physical quantities with closed spacelike 2-surfaces
[11-25]). The definition of the quasi-local energy-momentum, however, is
far from being so obvious as for example the Bondi--Sachs
energy-mom\-ent\-um. There are several inequivalent proposals for
it [26], and it is not clear how they are related to each other. The usual
formalism to carry out the calculations in the spinorial constructions
[18-25] is the elegant GHP formalism [27-29]. Since however its form is
not covariant, the geometric content of the expressions is not always
obvious and a lot of experience is needed to `see' the geometric content.
The covariance of a formalism may help and suggest why and what to
calculate. Furthermore, since the 3 dimensional Sen connection seems to be
`natural' in the energy-momentum problems of gravity,
one might conjecture that the two dimensional version of the Sen
connection has also significance in general relativity [30]. Thus in the
present paper, which is intended to be the first of a four paper series,
we would like to develop such a covariant spinor formalism that might be
the 2 dimensional version of the usual Sen connection and in which the
various constructions can be compared. This formalism may help to find the
`most natural' quasi-local energy-momentum expression; or at least yields a
better understanding of the energy-momentum problems of general relativity
by giving new insight into the geometry of the spacelike 2-surfaces.\par
      In the first two sections we review the geometry of spacelike 2
dimensional submanifolds, introduce the 2 dimensional version of the Sen
operator and calculate its curvature and torsion. Then, in section 3,
we discuss the algebra of surface spinors and find a complex metric
$\gamma_{AB}$ on the space of spinors. $\gamma_{AB}$ will have fundamental
importance in what follows. In section 4 the spinor form of the 2
dimensional Sen operator will be discussed, and, in section 5, we extend
the 2 dimensional covariant differentiation to spinors. We will see that
the 2 dimensional Sen connection is not simply a copy of the 3 dimensional
one in one less dimensions. An important difference between the one and
two co-dimensional submanifolds is that while in the one co-dimensional
case the normal is uniquely determined, in the two co-dimensional case
there is a 1 parameter family of unit normals, and as a consequence of this
`boost gauge freedom' the 2 dimensional Sen (and spin) connection has
always a `normal' piece as well.\par
       In section 6 the irreducible chiral parts of the 2 dimensional Sen
operator will be determined. They are precisely the right and left handed
parts of the 2-surface twistor and the 2 dimensional Weyl--Sen--Witten
operators, where the chirality is defined by the $\gamma^A{}_B$ spinor. In
section 7 we discuss the 2 dimensional counterparts of the 3 dimensional
Sen--Witten type identities. These are identities between the 2
dimensional Weyl--Sen--Witten and 2 dimensional twistor operators and
the integrand of various quasi-local charge integrals of the curvature and
the torsion of the 2 dimensional Sen operator. The Tod formula for the
kinematical twistor is a special consequence of them.\par
      In the present paper we work out only the general formalism. This
formalism will be applied only in the forthcoming papers for the
quasi-local energy-mom\-ent\-um, the quasi-local characterization of {\it
pp}-wave spacetimes and the gravitational radiative modes. Throughout the
present paper the abstract index formalism [28] will be used unless
otherwise stated. \par \vskip 1truecm

{\lbf 1. Two dimensional spacelike submanifolds}\par \vskip 0.5truecm

\noindent
Let $(M,g)$ be a four dimensional Lorentzian geometry with signature $-2$
and $\$ $ be a two dimensional spacelike submanifold. Let $t^a$ and $v^a$
be timelike and spacelike unit normals to $\$ $ being orthogonal to each
other: $t^at_a=1$, $v^av_a=-1$ and $t^av_a=0$. (If $\$ $ is orientable and
an open neighbourhood of $\$ $ in $M$ is space and time orientable, which
will be assumed in the present paper, then $t^a$ and $v^a$ are globally
well defined.) $t^a$ and $v^a$ are, of course, not unique, there is the
gauge freedom

$$
\eqalign{{t^\prime}^a&=t^a\cosh u+v^a\sinh u\cr
         {v^\prime}^a&=t^a\sinh u+v^a\cosh u.\cr}\eqno(1.1)
$$
\noindent (1.1) will be called a boost gauge transformation. The
projection to the tangent spaces of $\$ $ is given by

$$
\Pi^a_b:=\delta^a_b-t^at_b+v^av_b.
\eqno(1.2)$$

\noindent A tensor $T^{a_1...a_r}_{b_1...b_s}$ is called a surface tensor
if $T^{a_1...a_r}_{b_1...b_s}=\Pi^{a_1}_{e_1}...\Pi^{f_s}_{b_s}T^{e_1...
e_r}_{f_1...f_s}$. Specially $q_{ab}:=$ $\Pi^e_a\Pi^f_bg_{ef}$ $=g_{ab}-
t_at_b+v_av_b$ is the induced (negative definite) metric and (in the
non-abstract index formalism) $d\$={1\over2}t^av^b\varepsilon_{abcd}$
$dx^c\wedge dx^d$ is the induced volume element on $\$ $. Obviously, $\Pi
^a_b$, $q_{ab}$, $d\$ $ are all boost gauge invariant. \par
      For any {\it surface} vector field $X^a$ let us define

$$
\delta_aX^b:=\Pi^e_a\Pi^b_f\nabla_eX^f.
\eqno(1.3)$$

\noindent Since $\delta_aq_{bc}=0$ and $(\delta_a\delta_b-\delta_b\delta_a)
\phi=0$ for any function $\phi$, $\delta_a $ is the unique torsion free
metric Levi--Civit\`a covariant differentation. This is also boost gauge
invariant. \par
        To characterize the extrinsic geometry of $\$ $ in $M$ certain boost
gauge dependent quantities have to be introduced:

$$
\eqalign{\tau_{ab}:&=\Pi^e_a\Pi^f_b\nabla_et_f \cr
          \nu_{ab}:&=\Pi^e_a\Pi^f_b\nabla_ev_f \cr}
\eqno(1.4)$$

\noindent are the (symmetric) extrinsic curvatures and let

$$
A_a:=\Pi^f_a\nabla_ft_ev^e.\eqno(1.5)
$$

\noindent Under a boost gauge transformation they transform as

$$\eqalign{{\tau^\prime}_{ab}&=\tau_{ab}\cosh u+\nu_{ab}\sinh u\cr
            {\nu^\prime}_{ab}&=\tau_{ab}\sinh u+\nu_{ab}\cosh u\cr
                 {A^\prime}_a&=A_a-\delta_a u .\cr}
\eqno(1.6)$$
\noindent If the curvature tensors are defined by $R^a{}_{bcd}X^b:=-(
\nabla_c\nabla_d-\nabla_d\nabla_c)X^a$ and ${}^{\$}R^a{}_{bcd}$ $X^b:=
-(\delta_c\delta_d-\delta_d\delta_c)X^a$ for surface vectors, respectively,
then

$$
\eqalign{R_{efkl}\Pi^e_a\Pi^f_b\Pi^k_c\Pi^l_d&={}^{\$}R_{abcd}+\tau_{ac}
          \tau_{bd}-\tau_{ad}\tau_{bc}-\nu_{ac}\nu_{bd}+\nu_{ad}\nu_{bc}\cr
             t^aR_{afkl}\Pi^f_b\Pi^k_c\Pi^l_d&=\delta_c\tau_{db}-\delta_d
          \tau_{cb}+A_c\nu_{db}-A_d\nu_{cb}\cr
             v^aR_{afkl}\Pi^f_b\Pi^k_c\Pi^l_d&=\delta_c\nu_{db}-\delta_d
          \nu_{cb}+A_c\tau_{db}-A_d\tau_{cb}\cr
                 t^av^bR_{abkl}\Pi^k_c\Pi^l_d&=\tau_{ec}\nu^e{}_d-\tau_{ed}
            \nu^e{}_c+\delta_cA_d-\delta_dA_c.\cr}
\eqno(1.7)$$

\noindent The remaining six `irreducible' parts of the curvature, $t^at^c
R_{aecf}\Pi^e_b\Pi^f_d$, ...,$t^av^bt^cv^d$ $R_{abcd}$, can also be
expressed by the extrinsic curvatures, the vector potential $A_a$ and
additional boost gauge dependent quantities. However, they are rather
complicated and we do not need them. Since $\$ $ has dimension $2$, its
curvature tensor can be characterized completely by its curvature scalar:
${}^{\$}R_{abcd}={1\over2}{}^{\$}R(q_{ac}q_{bd}-q_{ad}q_{bc})$. \par
\vfill \eject

{\lbf 2. The two dimensional Sen operator $\Delta_a$}\par \vskip 0.5truecm

\noindent
Let us define the 2 dimensional version of the Sen operator: $\Delta_a:=
\Pi^e_a\nabla_e$. Obviously $\Delta_a$ is a differential operator acting
on {\it any} tensor field and annihilates the metric: $\Delta_ag_{bc}=0$.
Let

$$
Q^e{}_{ab}:=-\Pi^e_r\Delta_a\Pi^r_b=\tau^e{}_at_b-\nu^e{}_av_b. \eqno(2.1)
$$

\noindent This is by definition boost gauge invariant, and for any {\it
surface} vector $X^a$ one has

$$
\delta_aX_b=\Delta_aX_b+Q^e{}_{ab}X_e.\eqno(2.2)
$$

\noindent The action of the commutator of the $\Delta_a$'s on arbitrary
functions and vector fields are

$$
\eqalignno{(\Delta_a\Delta_b-\Delta_b\Delta_a)\phi&=-2Q^e{}_{[ab]}
                 \Delta_e\phi &(2.3)\cr
           (\Delta_a\Delta_b-\Delta_b\Delta_a)X^e &=-R^e{}_{frs}
	          \Pi^r_a\Pi^s_bX^f-2Q^f{}_{[ab]}\Delta_fX^e. &(2.4)\cr}
$$

\noindent The curvature and the torsion of $\Delta_a$ are therefore $F^e{}
_{fab}:=R^e{}_{frs}\Pi^r_a\Pi^s_b$ and $T^e{}_{ab}:=2Q^e{}_{[ab]}$,
respectively. By (2.1) the torsion is built up only from extrinsic
quantities, while the curvature can be expressed by the intrinsic geometry
of $\$ $, the extrinsic curvatures and the vector potential $A_a$
(eq.(1.7)). The curvature can be reexpressed by the intrinsic curvature,
the boost gauge independent $Q^e{}_{ab}$ and its $\Delta_a$--derivatives
and the field strength $\delta_cA_d-\delta_dA_c$ of the $A_c$ field:

$$
\eqalign{F_{abcd}={}^{\$}R_{abcd} &+\bigl(v_at_b-v_bt_a\bigr)
                             \Bigr(\delta_cA_d-\delta_dA_c\Bigr)+\cr
   &+2\Delta_dQ_{c[ab]}-2\Delta_cQ_{d[ab]}-4Q_{e[ab]}Q^e{}_{[cd]}+\cr
   &+g^{ef}\Bigl(Q_{ecb}Q_{fda}-Q_{edb}Q_{fca}-Q_{ace}Q_{bdf}+Q_{ade}Q_{bcf}
   \Bigr).\cr}
\eqno(2.5)$$

\noindent It might be worth noting that $\Delta _a$ is a covariant
differentiation in the sense of [31] on the pull back to $\$ $ of the tensor
bundle over $M$; and its curvature, defined by $-F^a{}_{bcd}X^bV^cZ^d :=
\Delta_V\Delta_ZX^a-\Delta_Z\Delta_VX^a-\Delta_{[V,Z]}X^a$ for any $X^a$
and surface vector fields $V^a$, $Z^a$, is precisely what we called
curvature. For connections on principle fibre bundles not isomorphic to a
(not necessarily nontrivial) reduced subbundle of the linear frame bundle
of the base manifold the torsion is not defined. Here the principle fibre
bundle on which the 2 dimensional Sen connection is defined is the pull
back to $\$ $ of the linear frame bundle $L(M)$. Thus the torsion of the
Sen connection cannot be defined in the strict sense of [31]. In fact,
while the curvature $F^a{}_{bcd}$ is a $gl(4,{\bf R})$ valued 2-form on
$\$ $, the `torsion' defined in (2.3) is {\it not} an ${\bf R}^4$ valued
2-form {\it on} $\$ $. However, in the calculations and formulae, e.g. in
(2.3,4), $2Q^e{}_{[ab]}$ behaves as a true torsion. This is the reason why
we will call $2Q^e{}_{[ab]}$ the torsion further on. \par
\vskip 1truecm

{\lbf 3. The algebra of 2-surface spinors}\par \vskip 0.5truecm

\noindent
If $t^{AA^\prime}$ and $v^{AA^\prime}$ are the spinor form of the normals to
$\$ $ then

$$
2t^{AR^\prime}t_{BR^\prime}=\delta^A_B, \hskip 1truecm 2v^{AR^\prime}
v_{BR^\prime}=-\delta^A_B \eqno(3.1)
$$

\noindent and let us define

$$
\gamma^A{}_B:=2t^{AR^\prime}v_{BR^\prime}. \eqno(3.2)
$$

\noindent It is easy to show that $\gamma^A{}_B$ is boost gauge
independent, invariant with respect to the conformal rescaling of the
metric and

$$
\gamma^R{}_R=0, \hskip 1truecm \gamma^A{}_R\gamma^R{}_B=\delta^A_B.
\eqno(3.3)$$

\noindent Thus $\gamma^A{}_B$ is nondegenerate and $\gamma:{\bf S}^A
\rightarrow{\bf S}^A:\lambda^A\mapsto\gamma^A{}_R\lambda^R$ is an
isomorphism of the spinor space ${\bf S}^A$ onto itself. Its eigenvalues
are $\pm1$, and hence $\gamma^A{}_B$ plays a role similar to the $\gamma
_5$ matrix. Its eigenspinors may be called left handed and right handed
with respect to $\gamma^A{}_B$. The spinor

$$
\pi^{\pm A}{}_B:={1\over2}(\delta^A_B\pm\gamma^A{}_B)\eqno(3.4)
$$

\noindent is the projection of the spin space to the subspace of
right/left handed spinors: $\gamma^A{}_R\pi^{\pm R}{}_B=\pm\pi^{\pm A}
{}_B$. The right/left handed spinors are pure spinors in the sense of [28].
(I am greatful to prof. H. Urbantke for this remark.) Right handed {\it
covariant} spinors should be defined by $\pi^{-R}{}_S$: $\mu_R=\mu_S\pi
^{-S}{}_R$. This definition ensures that $\mu_S$ is a right/left handed
covariant spinor iff $\mu^R$ is a right/left handed contravariant spinor.
The spinor form of the induced volume form on $\$ $ and on the 2-surface
element orthogonal to $\$ $ are given by

$$
\eqalign{t^av^b\varepsilon_{abcd}&={i\over2}\Bigl(\gamma_{CD}
     \varepsilon_{C^\prime D^\prime}-{\bar\gamma}_{C^\prime D^\prime}
     \varepsilon_{CD}\Bigr),\cr
 t_cv_d-t_dv_c&=-{1\over2}\Bigl(\gamma_{CD}\varepsilon_{C^\prime
      D^\prime}+\bar\gamma_{C^\prime D^\prime}\varepsilon_{CD}\Bigr),\cr}
\eqno(3.5)$$

\noindent
respectively. Thus geometrically $\gamma_{AB}$ is the anti-self-dual part
of the 2-surface element orthogonal to $\$ $. $\gamma_{AB}$ can also be
considered as a complex metric on ${\bf S}^A$. The null spinors of the
metric $\gamma_{AB}$ are just the eigenspinors of $\gamma^A{}_B$ and
hence $(\lambda^A,\gamma^A{}_R\lambda^R)$ is an independent system iff
$\lambda^A$ is not a null spinor. The group leaving invariant the complex
metric $\gamma_{AB}$ is isomorphic to ${\bf Z}_2\times{\bf C}^*$, where
${\bf C}^*:={\bf C}-\{0\}$; and the group leaving invariant both the
symplectic and complex metrics is ${\bf C}^*$. As a consequence of the
existence of the extra structure $\gamma_{AB}$ on ${\bf S}^A$ the
decomposition $\phi_{AB}={1\over2}\varepsilon_{AB}\phi_R{}^R$ $+\phi _{
(AB)}$ of a spinor $\phi_{AB}$ is not irreducible any more. Its symmetric
part can be decomposed further as $\phi_{(AB)}=-{1\over2}\gamma_{AB}
\gamma^{RS}\phi_{RS}$ $+\Bigl(\phi_{(AB)}+{1\over2}\gamma_{AB}\gamma^{RS}
\phi_{RS}\Bigr)$, the sum of the $\gamma$-trace and the trace-free
symmetric part of $\phi_{AB}$. The elements of the spinor space ${\bf
S}^A$ will be called spacelike 2 dimensional spinors if a spinor
$\gamma^A{}_B$ satisfying (3.3) is given on ${\bf S}^A$. (These `surface'
spinors should not be confused with the one component reduced spinors [28]
of the 2 dimensional geometry $(\$,q_{ab})$, which may also be called
surface spinors.)\par
       Let $(o_A,\iota_A)$ be a normalized spinor dyad such that

$$
t^a={1\over{\sqrt{2}}}\Bigl(o^A{\bar o}^{A^\prime}+\iota^A{\bar\iota}
            ^{A^\prime}\Bigr),  \hskip 1truecm
v^a={1\over{\sqrt{2}}}\Bigl(o^A{\bar o}^{A^\prime}-\iota^A{\bar\iota}
            ^{A^\prime}\Bigr). \eqno(3.6)
$$

\noindent Then $\gamma^A{}_B=o^A\iota_B+\iota^Ao_B$ and hence

$$
\gamma^A{}_Bo^B=-o^A, \hskip 1truecm
\gamma^A{}_B\iota^B=\iota^A.\eqno(3.7)
$$

\noindent Thus $o^A$ and $\iota^A$ are null spinors, $o^A$ is left handed
and $\iota^A$ is right handed. Conversely, if $o_A$, $\iota_A$ are left
and right handed spinors, respectively, satisfying $o_A\iota^A=1$ then they
form a GHP spinor dyad adapted to $\$ $. The linear isomorphism $\gamma:
{\bf S}^A\rightarrow{\bf S}^A$ can obviously be extended to the whole
tensor algebra over ${\bf S}^A$. Its action on the vectors of the complex
null tetrad is

$$
\eqalign{\gamma(l^a)&=\gamma(o^A){\bar\gamma}({\bar o}^{A^\prime})=l^a \cr
   \gamma(n^a)&=\gamma(\iota^A){\bar\gamma}({\bar\iota}^{A^\prime})=n^a\cr
   \gamma(m^a)&=\gamma(o^A){\bar\gamma}({\bar\iota}^{A^\prime})=-m^a.\cr}
\eqno(3.8)$$

\noindent Therefore a vector $X^a$ is tangent to $\$ $ iff $\gamma(X^a)=
-X^a$ and $X^a$ is orthogonal to $\$ $ iff $\gamma(X^a)=X^a$. The spinor
form of the projection $\Pi^a_b$ is therefore

$$
\Pi^a_b={1\over2}\Bigl(\delta^A_B\delta^{A^\prime}_{B^\prime}-\gamma^A{}_B
{\bar\gamma}^{A^\prime}{}_{B^\prime }\Bigr),\eqno(3.9)
$$

\noindent and hence $q_{AA^\prime BB^\prime}=$ ${1\over2}(\varepsilon_{AB}
\varepsilon_{A^\prime B^\prime}-$ $\gamma_{AB}{\bar\gamma}_{A^\prime
B^\prime})$. The vectors $l^a$, $n^a$, $m^a$ and $\bar m^a$ can also be
characterized as the coker of the projections $\pi^{-A}{}_B\bar\pi^{
-A^\prime}{}_{B^\prime}$, $\pi^{+A}{}_B\bar\pi^{+A^\prime}{}_{B^\prime}$,
$\pi^{-A}{}_B\bar\pi^{+A^\prime}{}_{B^\prime}$ and $\pi^{+A}{}_B\bar\pi^{
-A^\prime}{}_{B^\prime}$, respectively. $\pi^{\mp a}{}_b:=\pi^{\mp A}{}_B
\bar\pi^{\pm A^\prime}{}_{B^\prime}$ define a complex structure on the
tangent spaces of $\$ $ and they are the projections to the subspace of
(1,0) and (0,1) type vectors, respectively [32].\par
      For any spinor $\lambda^A$ let us define

$$
\eqalign{L^a&:=\bigl(\pi^{-A}{}_B\lambda^B\bigr)\bigl(\bar\pi^{-A^\prime}
     {}_{B^\prime}\bar\lambda^{B^\prime}\bigr)\cr
         N^a&:=\bigl(\pi^{+A}{}_B\lambda^B\bigr)\bigl(\bar\pi^{+A^\prime}
     {}_{B^\prime}\bar\lambda^{B^\prime}\bigr)\cr
         M^a&:=\bigl(\pi^{-A}{}_B\lambda^B\bigr)\bigl(\bar\pi^{+A^\prime}
     {}_{B^\prime}\bar\lambda^{B^\prime}\bigr). \cr}
\eqno(3.10)$$

\noindent It is easy to see that $L^a$ and $N^a$ are future directed null
vectors orthogonal to $\$ $, $M^a$ and $\bar M^a$ are complex null vectors
tangent to $\$ $ and $L^aN_a=-M^a{\bar M}_a=$ ${1\over4}\mid\gamma_{AB}
\lambda^A\lambda^B\mid^2$ and $\varepsilon_{abcd}L^aN^bM^c{\bar M}^d =
-{1\over16}\mid\gamma_{AB}\lambda^A\lambda^B\mid^4$. $\{L^a,N^a,M^a,{\bar
M}^a\}$ is therefore a future directed right handed complex null tetrad
adapted to $\$ $ unless $\lambda^A$ is a null spinor. If $\lambda^A$
becomes null then both $M^a$, ${\bar M}^a$ and either $L^a$ or $N^a$ will
be zero. The corresponding orthogonal vector basis is

$$
\eqalign{T^a&={1\over2\sqrt2}\Bigl(\lambda^A{\bar\lambda}^{A^\prime}+
     \gamma^A{}_R\lambda^R{\bar\gamma}^{A^\prime}{}_{R^\prime}{\bar
     \lambda}^{R^\prime}\Bigr)   \hskip 1truecm
Z^a=-{1\over2\sqrt2}\Bigl(\lambda^A{\bar\gamma}^{A^\prime}{}_{R^\prime}
     {\bar\lambda}^{R^\prime}+\gamma^A{}_R\lambda^R{\bar\lambda}^{A^\prime}
     \Bigr)\cr
X^a&={1\over2\sqrt2}\Bigl(\lambda^A{\bar\lambda}^{A^\prime}-\gamma^A{}_R
     \lambda^R{\bar\gamma}^{A^\prime}{}_{R^\prime}{\bar\lambda}^{R^\prime}
     \Bigr)              \hskip 1truecm
Y^a=-{i\over2\sqrt2}\Bigl(\gamma^A{}_R\lambda^R{\bar\lambda}^{R^\prime}-
     \lambda^A{\bar\gamma}^{A^\prime}{}_{R^\prime}{\bar\lambda}^{R^\prime}
     \Bigr).\cr} \eqno(3.11)
$$

\noindent The length of these vectors is ${1\over4}\mid\gamma_{AB}
\lambda^A\lambda^B\mid^2$. If $\lambda^A$ becomes null then $T^a$ and
$Z^a$ will be parallel null vectors. Thus a single non-null spinor field
on $\$ $ is able to define an orthogonal vector basis in the Lorentzian
tangent spaces. Similar statement holds for any nonzero spinor field on a
spacelike hypersurface [33].\par
     The spinor $\gamma^A{}_B$ can be defined by intrinsic quantities too:
if $x^{AA^\prime}$ and $y^{AA^\prime}$ are $q_{ab}$-orthonormal vectors
tangent to $\$ $ then $J^A{}_B:=2x^{AR^\prime}y_{BR^\prime}=i\gamma^A{}_B$.
$J^A{}_B$, and hence $\gamma^A{}_B$ also, is rotation gauge invariant. The
fact that $\gamma^A{}_B$ is globally well defined is a consequence of its
definition (3.2) and the globality of $t^a$ and $v^a$. If we defined
$\gamma^A{}_B$ by the intrinsic properties of $\$ $ then it would not {\it
by definition} be globally well defined. But, using its rotation gauge
invariance, it would be easy to show that it is, in fact, globally well
defined. \par
\vfill \eject

{\lbf 4. The spinor form of $\Delta _a$}\par \vskip 0.5truecm

\noindent
The action of the commutator of the $\Delta_a$'s on a spinor field:

$$
\Bigl(\Delta_c\Delta_d-\Delta_d\Delta_c\Bigr)\xi^A=
-R^A{}_{Bef}\Pi^e_c\Pi^f_d\xi^B-2Q^e{}_{[cd]}\Delta_e\xi^A,\eqno(4.1)
$$

\noindent where $R^A{}_{Bef}:={1\over2}R^{AA^\prime }{}_{BA^\prime ef}$ is
the anti-self-dual part of the spacetime curvature, and hence the
curvature of the operator $\Delta_c$ is the pull back to $\$ $
of the anti-self-dual part of the spacetime curvature. In terms of the Weyl
and Ricci spinors and the $\Lambda$ scalar it is given by

$$
\eqalign{F_{ABCC^\prime DD^\prime}:=R_{ABef}&\Pi^e_{CC^\prime}\Pi^f
_{DD^\prime}=\cr
=-{1\over4}\varepsilon_{C^\prime D^\prime} &\Bigl(\psi_{ABCD}-\psi_{ABEF}
\gamma^E{}_C\gamma^F{}_D+\gamma_{CD}\phi_{ABE^\prime F^\prime}{\bar
\gamma}^{E^\prime F^\prime}+\cr &+\Lambda\bigl(\varepsilon_{AC}\varepsilon
_{BD}+\varepsilon_{AD}\varepsilon_{BC}-\gamma_{AC}\gamma_{BD}-\gamma
_{AD}\gamma_{BC}\bigr)\Bigr)-\cr
-{1\over4}\varepsilon_{CD} &\Bigl({\bar\gamma}_{C^\prime D^\prime}\psi
_{ABEF}\gamma^{EF}+\phi_{ABC^\prime D^\prime}-\phi_{ABE^\prime F^\prime}
{\bar\gamma}^{E^\prime}{}_{C^\prime}{\bar\gamma}^{F^\prime}{}_{D^\prime}+
\cr
&+2\Lambda\gamma_{AB}{\bar\gamma}_{C^\prime D^\prime}\Bigr).\cr}
\eqno(4.2)$$

\noindent Its contraction with the induced volume form:

$$
F_{ABcd}t^ev^f\varepsilon_{ef}{}^{cd}=-i\Bigl(\psi_{ABCD}\gamma^{CD}-
\phi_{ABC^\prime D^\prime}{\bar\gamma}^{C^\prime D^\prime}+2\Lambda\gamma
_{AB} \Bigr).\eqno(4.3)
$$

To find the spinor form of the torsion and the expression of the curvature
in terms of extrinsic and intrinsic geometrical quantities, let us define

$$
Q^E{}_{aF}:={1\over2}\Delta_a\gamma^E{}_R\gamma^R{}_F.\eqno(4.4)
$$

\noindent Then obviously $Q^R{}_{aR}=0$ and by the definition (2.1) of
$Q^e{}_{ab}$ and the spinor form (3.9) of the projection one has

$$
Q^e{}_{ab}={1\over2}\Bigl(\delta^{E^\prime}_{B^\prime}Q^E{}_{aB}+\delta
^E _B {\bar Q}^{E^\prime}{}_{aB^\prime}+Q^E{}_{aR}\gamma^R{}_B{\bar\gamma
}^{E^\prime}{}_{B^\prime}+{\bar Q}^{E^\prime}{}_{aR^\prime}{\bar\gamma
}^{R^\prime}{}_{B^\prime}\gamma^E{}_B\Bigr).\eqno(4.5)
$$

\noindent Since $Q_{eab}=Q_{aeb}$ (cf. eq.(2.1)), $Q^E{}_{aF}$ has the
`hidden' symmetry

$$
Q_{EAA^\prime F}={1\over2}\Bigl(\delta^{R^\prime}_{A^\prime}\delta^R_F
-{\bar\gamma}^{R^\prime}{}_{A^\prime}\gamma^R{}_F\Bigr)Q_{AER^\prime R}+
{1\over2}\Bigl(\varepsilon_{AF}\varepsilon^{E^\prime F^\prime}-\gamma
_{AF}{\bar\gamma}^{E^\prime F^\prime}\Bigr){\bar Q}_{A^\prime EE^\prime
F^\prime}.\eqno(4.6)
$$

\noindent The spinor form of the torsion is therefore

$$
T_{EE^\prime AA^\prime BB^\prime}=-\Bigl(\varepsilon_{A^\prime B^\prime}
Q_{AEE^\prime B}+\varepsilon_{AB}{\bar Q}_{A^\prime EE^\prime B^\prime
}\Bigr), \eqno(4.7)
$$

\noindent and the expression (2.5) for the curvature is

$$
\eqalign{F^A{}_{Bcd}&={1\over2}{}^{\$}R^{AA^\prime}{}_{BA^\prime cd}
-{1\over2}\gamma^A{}_B\Bigl(\delta_c A_d-\delta_dA_c\Bigr)+\cr
&+\Delta_{DD^\prime}Q^A{}_{CC^\prime B}-\Delta_{CC^\prime}Q^A{}_{DD^\prime
B}+Q^A{}_{CC^\prime R}Q^R{}_{DD^\prime B}-Q^A{}_{DD^\prime R}Q^R{}_{C
C^\prime B}+\cr
&+Q^A{}_{RR^\prime B}\Bigl(\delta^{R^\prime}_{C^\prime}Q^R{}_{DD^\prime C}+
\delta^R_C\bar Q^{R^\prime}{}_{DD^\prime C^\prime}-\delta^{R^\prime}_{
D^\prime}Q^R{}_{CC^\prime D}-\delta^R_D\bar Q^{R^\prime}{}_{CC^\prime
D^\prime}\Bigr);\cr}\eqno(4.8)
$$

\noindent where ${}^{\$}R_{AA^\prime BB^\prime cd}$ is the curvature tensor
of $\$ $. As one may expect, the torsion contains all the information on the
divergences and shears of the null geodesics orthogonal to $\$ $. In fact,

$$
\eqalign{o^Ao^Bo^R{\bar\iota}^{R^\prime}Q_{ARR^\prime B}&=\sigma
  \hskip 15pt
  \iota^A\iota^Bo^R{\bar\iota}^{R^\prime}Q_{ARR^\prime B}=-\rho^\prime \cr
o^Ao^B\iota^R{\bar o}^{R^\prime}Q_{ARR^\prime B}&=\rho \hskip 15pt
  \iota^A\iota^B\iota^R{\bar o}^{R^\prime}Q_{ARR^\prime B}=-\sigma^\prime
  \cr}\eqno(4.9)
$$

\noindent are the familiar GHP spin coeffitients and the remaining
contractions are zero. As a consequence of the `hidden' symmetry (4.6)
$\rho$ and $\rho^\prime$ are, of course, real.\par
\vskip 1truecm

{\lbf 5. The induced 2-surface spin connection}\par \vskip 0.5truecm

\noindent
To motivate how to define the action of $\delta_a$ on spinors let us
consider the $\delta_a$--derivative of the complex null surface vector
$M^a$ defined by eq.(3.10):

$$
\eqalign{\delta_aM^b&=\Delta_aM^b-\Delta_a\Pi^b_fM^f=\cr
      &={1\over4}\Bigl(\Delta_a(\lambda^B-\gamma^B{}_R\lambda^R)+
        {1\over2}\Delta_a\gamma^B{}_R(\lambda^R-\gamma^R{}_S\lambda^S)\Bigr)
	 ({\bar\lambda}^{B^\prime}+{\bar\gamma}^{B^\prime}{}_{R^\prime}{\bar
	 \lambda}^{R^\prime})+\cr
      &+{1\over4}(\lambda^B-\gamma^B{}_R\lambda^R)\Bigl(\Delta_a({\bar
        \lambda}^{B^\prime}+{\bar\gamma}^{B^\prime}{}_{R^\prime}{\bar
	 \lambda}^{R^\prime})-{1\over2}\Delta_a{\bar\gamma}^{B^\prime}
	 {}_{R^\prime}({\bar\lambda}^{R^\prime}+{\bar\gamma}^{B^\prime}
	 {}_{S^\prime}{\bar\lambda}^{S^\prime})\Bigr).\cr}
\eqno(5.1)$$

\noindent Thus it seems natural to define the action of $\delta_a$ on
spinors by

$$
\delta_a(\lambda^B\pm\gamma^B{}_R\lambda^R):=\Delta_a(\lambda^B\pm
\gamma^B{}_R\lambda^R)\mp{1\over2}\Delta_a\gamma^B{}_R(\lambda^R\pm
\gamma^R{}_S\lambda^S);
$$

\noindent i.e. by

$$
\delta_a\lambda^B:=\Delta_a\lambda^B-Q^B{}_{aR}\lambda^R.\eqno(5.2)
$$

\noindent $\delta_a$ annihilates both the symplectic and complex metrics:

$$
\delta_a\varepsilon_{RS}=0, \hskip 1truecm \delta_a\gamma_{RS}=0.\eqno(5.3)
$$

\noindent Therefore the 2-surface spinor curvature, defined by

$$
{}^{\$}R^A{}_{Bcd}\xi^B:=-(\delta_c\delta_d-\delta_d\delta_c )\xi^A,
\eqno(5.4)$$

\noindent has the algebraic symmetries ${}^{\$}R_{ABcd}=$ ${}^{\$}R
_{BAcd}$ and $\gamma_{AD}{}^{\$}R^D{}_{Bcd}$ $={1\over2}\varepsilon_{AB}
\gamma^{EF}{}^{\$}R_{EFcd};$ and hence

$$
{}^{\$}R_{ABcd}=-{1\over2}\gamma_{AB}\gamma^{EF}{}^{\$}R_{EFcd}.
\eqno(5.5)$$

\noindent Calculating the action of the commutator $(\delta_c\delta_d
-\delta_d\delta_c)$ on the surface vector $\Pi^{AA^\prime}_{BB^\prime}
\lambda^B\bar\lambda^{B^\prime}$ we obtain the spinor form of the
curvature tensor of $\$ $ by the spinor curvature:

$$
{}^{\$}R_{AA^\prime BB^\prime cd}={i\over2}\bigl(\varepsilon_{A^\prime
B^\prime}\gamma_{AB}-\varepsilon_{AB}\bar\gamma_{A^\prime B^\prime}\bigr)
{i\over2}\Bigl(-\gamma^{EF}{}^{\$}R_{EFcd}+\bar\gamma^{E^\prime F^\prime}
{}^{\$}\bar R_{E^\prime F^\prime cd}\Bigr).\eqno(5.6)
$$

\noindent This is the product of the surface volume form and the imaginary
part of $\gamma^{AB}{}^{\$}R_{ABcd}$, and hence, contracting with the
volume form and using the expression of the curvature tensor ${}^{\$}
R_{abcd}$ by the induced metric and the curvature scalar ${}^{\$}R$, the
imaginary part of $\gamma^{AB}{}^{\$}R_{ABcd}$ can be expressed by the
curvature scalar ${}^{\$}R$ and the volume form. To determine the full
spinor curvature first let us rewrite the commutator $(\delta_c\delta_d-
\delta_d\delta_c)\xi^A$ by (5.2) and (4.1) to obtain

$$
\eqalign{-{}^{\$}R_{ABcd}&=-F_{ABcd}+\Delta_{DD^\prime}Q_{ACC^\prime B}
-\Delta_{CC^\prime}Q_{ADD^\prime B}+Q_{ACC^\prime R}Q^R{}_{DD^\prime B}
-Q_{ADD^\prime R}Q^R{}_{CC^\prime B}+\cr
&+Q_{ARR^\prime B}\Bigl(\delta^{R^\prime}_{C^\prime}Q^R{}_{DD^\prime C}
+\delta^R_C\bar Q^{R^\prime}{}_{DD^\prime C^\prime}-\delta^{R^\prime}
_{D^\prime}Q^R{}_{CC^\prime D}-\delta^R_D\bar Q^{R^\prime}{}_{CC^\prime
D^\prime}\Bigr).\cr}\eqno(5.7)
$$

\noindent Comparing its right hand side with (4.8) we have

$$
{}^{\$}R_{ABcd}=-{1\over2}\gamma_{AB}\Bigl(\bigl(\delta_cA_d-\delta_d
A_c\bigr)-{{}^{\$}R\over4}\bigl(\varepsilon_{C^\prime D^\prime }\gamma
_{CD}-\varepsilon_{CD}\bar\gamma_{C^\prime D^\prime}\bigr)\Bigr).\eqno(5.8)
$$

\noindent Thus the imaginary part of $\gamma^{AB}{}^{\$}R_{ABcd}$ is, in
fact, related to the curvature tensor ${}^{\$}R_{abcd}$; i.e. to the
intrinsic geometry of $\$ $. However, its real part is an extrinsic
geometrical quantity: the `field strength' of the vector potential $A_e$.
Thus the spinor curvature is {\it not} the anti-self-dual part of the
intrinsic curvature of $\$ $. With this extension of $\delta_e$ from
surface tensors to spinors we have extended $\delta_e$ to arbitrary
tensors: For any vector field $X^a$ orthogonal to $\$ $ $\delta_eX^a=\Pi^f
_e(\delta^a_b-\Pi^a_b)\nabla_fX^b$. If $(o^R$, $\iota^R)$ is a spinor dyad
normalized by (3.6) then twice the real part of the spinor connection
coeffitient, $B_e:=-o^A\Delta_e\iota_A$, is the $SO(1,1)$-gauge potential:
$B_e+\bar B_e=A_e$, while the only independent part of the $SO(2)$-Ricci
rotation coeffitients is $B_e-\bar B_e=\bar m^a\delta_em_a$. The GHP spin
coeffitients representing the connection are related to $B_e$ by
$\beta=-B_em^e$ and $\beta^\prime=B_e\bar m^e$, respestively. The
curvature associated with the action of $\delta_e$ on vectors orthogonal
to $\$ $ is ${1\over2}(\varepsilon_{A^\prime B^\prime}\gamma_{AB}+\varepsilon
_{AB}\bar\gamma_{A^\prime B^\prime})(\delta_cA_d-\delta_dA_c)$. This result
is in accordance with the following geometrical picture [31]: The surface
$\$ $ defines a principle fibre bundle over $\$ $ with structure group
$SO(2)\otimes SO(1,1)$ such that $SO(2)$ acts naturally on the tangent
bundle and $SO(1,1)$ on the normal bundle of $\$ $. The principle ${\bf
C}^*$-bundle $(B,\$,{\bf C}^*)$ of the normalized spinor dyads satisfying
(3.6) over $\$ $ [34] is its double covering bundle. Since this principle
${\bf C}^*\approx U(1)\otimes(0,\infty)$-bundle over $\$ $ is the pull back
to $\$ $ of the sum of the principle $U(1)$- and $(0,\infty)$-bundles along
the diagonal map $\$\rightarrow\$\times\$: $ $p\mapsto (p,p)$, the
curvature of any connection on $(B,\$,{\bf C}^*)$ must be the sum of the
two curvatures corresponding to the $U(1)$ and $(0,\infty)$ subgroups,
respectively. In fact, the real part of the spinor curvature is connected
with the $(0,\infty)$, while its imaginary part with the $U(1)$ subgroup.
\par
        Finally contracting (5.8) with $\gamma^{AB}$ and integrating for
$\$ $:

$$
\eqalign{\int_{\$}\gamma^{AB}{}^{\$}R_{ABcd}dx^c\wedge dx^d&=
2\int_{\$}dA+i\int_{\$}{}^{\$}R{1\over2}t^av^b\varepsilon_{abcd}
dx^c\wedge dx^d=\cr
&=2\oint_{\partial\$}A+i\int_{\$}{}^{\$}Rd\$.\cr}\eqno(5.9)
$$

\noindent Thus for closed $\$ $ the real part of ${}^{\$}R_{ABcd}$ does not
contribute to the total, integral curvature; and by the Gauss--Bonnet
theorem the total curvature is $8\pi i(1-{\cal G})$ where ${\cal G}$ is
the genus of $\$ $. This is a simple, direct verification of the fact [28]
that the integral for a closed $\$ $ of the imaginary part of the complex
Gauss curvature of $\$ $, given by $K=-{i\over2}t^av^b\varepsilon_{ab}
{}^{cd}\delta_cA_d+{1\over4}{}^{\$}R$ in the present formalism, is zero.\par
\vfill \eject

{\lbf 6. The irreducible parts of $\Delta_b$}\par \vskip 0.5truecm

\noindent There are essentially two irreducible parts of the first
Sen--derivative of a spinor field: the contraction $\Delta_{R^\prime}{}^R
\lambda_R$ and

$$
{\cal T}_{R^\prime RS}{}^K\lambda_K:=\Delta_{R^\prime(R}\lambda_{S)}+
{1\over2}\gamma_{RS}\gamma^{EF}\Delta_{R^\prime E}\lambda_F,\eqno(6.1)
$$

\noindent the trace-free symmetric part of $\Delta_{R^\prime R}\lambda_S$.
The `$\gamma$-trace' $\gamma^{RS}\Delta_{R^\prime R}\lambda_S$ is not
independent since, because of the argumentation following eq.(3.8), $\gamma
^{RS}\Delta_{R^\prime R}\lambda_S={\bar\gamma}_{R^\prime}{}^{S^\prime}
\Delta_{S^\prime R}\lambda^R$; and hence, using (3.9),

$$
\Delta_{RR^\prime}\lambda_S=\Pi^{A^\prime A}_{R^\prime R}\varepsilon_{AS}
\Delta_{A^\prime K}\lambda^K+{\cal T}_{R^\prime RS}{}^K\lambda_K.
\eqno(6.2)$$

\noindent $\Delta_{R^\prime}{}^R\lambda_R=0$ is the 2 dimensional version
of the Sen--Witten equation [5,9,10]. Recalling that under the conformal
rescaling $\varepsilon_{AB}\mapsto$ $\tilde\varepsilon_{AB}:=\Omega
\varepsilon_{AB}$ the spacetime covariant differentiation is known to
transform [28] as $\nabla_{RR^\prime}\psi^{A...}_{B...}\mapsto$ $\tilde
\nabla_{RR^\prime}\psi^{A...}_{B...}:=$ $\nabla_{RR^\prime}\psi^{A...}
_{B...}+$ $C^A{}_{RR^\prime E}\psi^{E...}_{B...}+...$ $-C^F{}_{RR^\prime B}
\psi^{A...}_{F...}-...$, where $C^E{}_{RR^\prime F}:=\delta^E_R\Upsilon_{F
R^\prime}$ and $\Upsilon_e:=\nabla_e\ln\Omega$, it is easy to deduce the
behaviour of $\Delta_{RR^\prime}$, $Q_{ARR^\prime B}$,.. etc. under
conformal rescalings. In particular if $\lambda_R$ has conformal weight
$w\in{\bf R}$ (i.e. $\lambda_R\mapsto\tilde\lambda_R:=\Omega^w\lambda_R$
under the conformal rescaling) then the conformal bahaviour of ${\cal T}_{
R^\prime RS}{}^K$ and the 2 dimensional Weyl--Sen--Witten operators:

$$
\eqalignno{\tilde\Delta_{R^\prime}{}^R\tilde\lambda_R&=\Omega^{w-1}\Delta
_{R^\prime}{}^R\lambda_R-\Omega^{w-1}\lambda^R{1\over2}\Bigl((1+w)\delta^K
_R\delta^{K^\prime}_{R^\prime}+(1-w)\gamma^K{}_R\bar\gamma^{K^\prime}{}_{
R^\prime}\Bigr)\Upsilon_{KK^\prime} &(6.3)\cr
\tilde{\cal T}_{R^\prime RS}{}^K\tilde\lambda_K&=\Omega^w {\cal T}_{
R^\prime RS}{}^K\lambda_K+{1\over4}\Omega^w\lambda_B\bigl(\delta^B_S\gamma
^K{}_R+\delta^K_S\gamma^B{}_R+\delta^B_R\gamma^K{}_S+\delta^K_R\gamma^B{}_S
\bigr)\bar\gamma^{K^\prime}{}_{R^\prime}\Upsilon_{KK^\prime}\cr
+{1\over4}&\Omega^w\bigl(\gamma_{RS}\lambda^B-\gamma^B{}_R\lambda_S-\gamma
^B{}_S\lambda_R\bigr)\Bigl((1+w)\delta^K_B\bar\gamma^{K^\prime}{}_{R^\prime}
+(1-w)\gamma^K{}_B\delta^{K^\prime}_{R^\prime}\Bigr)\Upsilon_{KK^\prime}.
&(6.4)\cr}$$

\noindent Thus, in contrast to the four dimensional Weyl
neutrino operator $\nabla_{R^\prime}{}^R$, the Weyl--Sen--Witten operator
$\Delta_{R^\prime}{}^R$ does not have definite conformal weight, while
${\cal T}_{R^\prime RS}{}^K$ has zero conformal weight if it acts on spinor
fields of unit conformal weight. However if $\lambda_R$ has unit conformal
weight then under the conformal rescaling the spinor

$$
\pi_{A^\prime}:=-i\Delta_{A^\prime}{}^A\lambda_A\eqno(6.5)
$$

\noindent transforms like the secondary part of a twistor; i.e. ${\tt Z}
^\alpha:=(\lambda^A,\pi_{A^\prime})$ is a local twistor [1] on $\$ $. One
can calculate the covariant derivative of a local twistor defined on
$\$ $ in the direction tangential to $\$ $; i.e. to define the 2
dimensional Sen derivative of any local twistor ${\tt Z}^\alpha=(\lambda^A,
\pi_{A^\prime})$ defined on $\$ $:

$$
\Delta_b{\tt Z}^\alpha:=\Pi^e_b\nabla_e{\tt Z}^\alpha=\Bigl(\Delta_{B
B^\prime}\lambda^A+i\Pi^{AA^\prime}_{BB^\prime}\pi_{A^\prime},\hskip 5pt
\Delta_{BB^\prime}\pi_{A^\prime}+i\lambda^A{1\over2}\bigl({1\over6}Rg_{ae}
-R_{ae}\bigr)\Pi^{EE^\prime}_{BB^\prime}\Bigr).\eqno(6.6)
$$

\noindent Its primary part is ${\cal T}_{B^\prime B}{}^{AK}\lambda_K+i\Pi
^{AK^\prime}_{BB^\prime}(\pi_{K^\prime}+i\Delta_{K^\prime}{}^K\lambda_K)$;
and hence the primary part of the Sen derivative of a twistor satisfying
(6.5) is just the ${\cal T}_{R^\prime RS}{}^K$-derivative of the primary
spinor part of the twistor. Thus ${\cal T}_{R^\prime RS}{}^K$ is precisely
the 2-surface twistor operator. (Borrowing the idea how the twistor
covariant differentiation, $\nabla_b{\tt Z}^\alpha$, is defined [1] one
can define the induced 2 dimensional covariant derivative $\delta_b{\tt
Z}^\alpha$ of ${\tt Z}^\alpha$ too. This, however, will not be used in the
present paper.) \par
        It will be useful to introduce the following chiral differential
operators:

$$
\eqalignno{\Delta^{\pm R}_{R^\prime}\lambda_R&:={\bar\pi}^{\mp S^\prime}
   {}_{R^\prime}\Delta_{S^\prime}{}^R\lambda_R, &(6.7)\cr
{\cal T}^{\pm}_{R^\prime RS}{}^K\lambda_K&:={\bar\pi}^{\mp S^\prime}
   {}_{R^\prime}{\cal T}_{S^\prime RS}{}^K\lambda_K.&(6.8)\cr}
$$

\noindent ${\cal T}^\pm_{R^\prime RS}{}^K$ may be called the right/left
handed parts of the twistor operator ${\cal T}_{R^\prime RS}{}^K$ and
$\Delta^{\pm R}_{R^\prime}$ the right/left handed part of the
Weyl--Sen--Witten operator. Any further application of the symmetry
operations and the projections on ${\cal T}^\pm_{A^\prime AB}{}^R$ and
$\Delta^\pm_{R^\prime}{}^R$ yields zero or is an identity; i.e. ${\cal
T}^\pm_{A^\prime AB}{}^K$, $\Delta_{R^\prime}^{\pm R}$ form the complete
irreducible decomposition of $\Delta_{A^\prime A}\lambda_B$. \par
     Finally determine the GHP form of these irreducible parts. Let $(o_A,
\iota_A)$ be a spinor dyad normalized by $o_A\iota^A=1$ and (3.6), and
define $\lambda^0=\lambda_1$ and $\lambda^1=-\lambda_0$ by $\lambda^A=:
\lambda^0o^A+\lambda^1\iota^A$. They are scalars of weight $(-1,0)$ and
$(1,0)$, respectively [27-29]. Then the GHP form of the irreducible chiral
operators are

$$
\eqalignno{-{\bar\iota}^{R^\prime}\Delta^{-R}_{R^\prime}\lambda_R &=
  \iota^Rm^a\nabla_a\lambda_R=\edth\lambda^0-\rho^\prime\lambda^1,&(6.9)\cr
-{\bar o}^{R^\prime}\Delta^{+R}_{R^\prime}\lambda_R&=-o^R{\bar m}^a\nabla_a
  \lambda_R=\edth^\prime\lambda^1-\rho\lambda^0,&(6.10)\cr
-{\bar\iota}^{R^\prime}o^Ro^S{\cal T}^-_{R^\prime RS}{}^K\lambda_K&=-o^Rm^a
  \nabla_a\lambda_R=\edth\lambda^1-\sigma\lambda^0, &(6.11)\cr
{\bar o}^{R^\prime}\iota^R\iota^S{\cal T}^+_{R^\prime RS}{}^K\lambda_K&=
  \iota^R{\bar m}^a\nabla_a\lambda_R=\edth^\prime\lambda^0-\sigma^\prime
  \lambda^1;&(6.12)\cr}
$$

\noindent and all the remaining contractions are zero. Thus the GHP form
of ${\cal T}_{R^\prime RS}{}^K\lambda_K$ $=0$ is the familiar 2-surface
twistor equation [1,18-20]. The GHP form (6.9-12) of the irreducible chiral
parts of the 2 dimensional Sen operator define differential operators on
the Whitney sum of certain vector bundles $E(p,q)$ of scalars of weight
$(p,q)$, $p-q\in{\bf Z}$.

$$
\eqalign{-\Delta^-:&E^\infty(p-1,q)\oplus E^\infty(p+1,q)\rightarrow
   E^\infty(p,q-1):(\lambda^0,\lambda^1)\mapsto\bigl(\edth\lambda^0-\rho
   ^\prime\lambda^1\bigr), \cr
-\Delta^+:&E^\infty(p-1,q)\oplus E^\infty(p+1,q)\rightarrow E^\infty(p,
   q+1):(\lambda^0,\lambda^1)\mapsto\bigl(\edth^\prime\lambda^1-\rho\lambda
   ^0\bigr), \cr
-{\cal T}^-:&E^\infty(p-1,q)\oplus E^\infty(p+1,q)\rightarrow E^\infty(p+2,
   q-1):(\lambda^0,\lambda^1)\mapsto\bigl(\edth\lambda^1-\sigma\lambda^0
   \bigr),\cr
-{\cal T}^+:&E^\infty(p-1,q)\oplus E^\infty(p+1,q)\rightarrow E^\infty(p-2,
   q+1):(\lambda^0,\lambda^1)\mapsto\bigl(\edth^\prime\lambda^0-\sigma
   ^\prime\lambda^1\bigr).\cr}
\eqno(6.13)$$

\noindent Here $E^\infty(p,q)$ is the space of the smooth cross sections
of $E(p,q)$ (see e.g. [34]). For $p=q=0$ these operators reduce to the
GHP form of the irreducible chiral parts of the 2 dimensional
Weyl--Sen--Witten and twistor operators acting on the space of the smooth
covariant spinor fields $C^\infty(\$,{\bf S}_A)\simeq E^\infty(-1,0)
\oplus E^\infty(1,0)$. \par \vskip 1truecm

{\lbf 7. The spinor identities}\par

\vskip 0.5truecm
\noindent Using (5.2), the commutator (4.1) and (4.2-3) one has the
following identity for any two spinor fields $\lambda^A$ and $\mu^A$:

$$
\eqalign{{1\over2}{\bar\gamma}^{A^\prime B^\prime}(\Delta_{A^\prime A}
\lambda^A)(\Delta_{B^\prime B}\mu^B)&=\delta_{AA^\prime}\Bigl({\bar\gamma}
^{A^\prime B^\prime}\lambda_B\Delta^A_{B^\prime}\mu^B\Bigr)-\cr
-{\bar\gamma}^{A^\prime B^\prime}(\Delta_{A^\prime(A}\lambda_{B)}) &(\Delta
_{B^\prime}^{(A}\mu^{B)})-{i\over2}\lambda^A\mu^BR_{ABcd}t^ev^f\varepsilon
_{ef}{}^{cd}.\cr}\eqno(7.1)
$$

\noindent This is the two dimensional version [30] of the Sen identity
[9,10]: apart from the total divergence, this is a relation between the
first derivatives of the spinor fields and the curvature, which is actually
the integrand of Penrose's quasi-local charge integrals. Instead of the
complex inner product of $\Delta_{A^\prime A}\lambda^A$ and $\Delta_{A
^\prime A}\mu^A$ one can start with their symplectic inner product. Then
using (5.2), (4.1-3) and (4.7) we have

$$
\eqalignno{{3\over2}\varepsilon^{A^\prime B^\prime}&(\Delta_{A^\prime A}
\mu^A)(\Delta_{B^\prime B}\lambda^B)=\delta_{RR^\prime}\Bigl(\mu^R\Delta^{
R^\prime}_S\lambda^S+\mu^S\Delta^{R^\prime}_S\lambda^R\Bigr)+\varepsilon
^{A^\prime B^\prime}(\Delta_{A^\prime (A}\mu_{B)})(\Delta_{B^\prime}^{(A}
\lambda^{B)})-\cr
&-{i\over2}\mu^A\gamma_A{}^K\lambda^BR_{KBcd}t^ev^f\varepsilon_{ef}{}^{cd}
-&(7.2)\cr
&-{1\over2}\Bigl(2\varepsilon^{RA}\mu^B\Delta^{R^\prime}_S\lambda^S-\mu^R
\Delta^{R^\prime(A}\lambda^{B)}-\mu^B\Delta^{R^\prime(A}\lambda^{R)}-
\varepsilon^{RA}\mu_S\Delta^{R^\prime(S}\lambda^{B)}\Bigr)\varepsilon^{
A^\prime B^\prime}T_{RR^\prime AA^\prime BB^\prime}.\cr}
$$

\noindent The second derivatives of the spinor fields appear only in the
form of a total divergence again [30]. However, in contrast to the identity
(7.1), eq.(7.2) contains the torsion $T_{rab}$ of the operator $\Delta_a$
too. If we integrate the identities (7.1) and (7.2) for the spacelike
2-submanifold $\$ $ which is orientable and {\it closed} then the total
divergences disappear. Expressing the $\Delta_a$ operators by the two
dimensional Weyl--Sen--Witten and twistor operators we obtain identities
between the integral of the Weyl--Sen--Witten and twistor derivatives and
the charge integrals of the curvature and the torsion of $\Delta_a$:

$$
\oint_{\$}{\bar\gamma}^{R^\prime S^\prime}\Bigl(\Delta_{R^\prime}{}^R
\lambda_R\Delta_{S^\prime}{}^S\mu_S+{\cal T}_{R^\prime RS}{}^K\lambda_K
{\cal T}_{S^\prime}{}^{RSL}\mu_L\Bigr)d\$=-{i\over2}\oint_{\$}\lambda^A\mu
^BR_{ABcd}dx^c\wedge dx^d\eqno(7.3)
$$

\noindent and

$$
\eqalign{\oint_{\$}\varepsilon^{R^\prime S^\prime}\Bigl(\Delta_{R^\prime}
{}^R\mu_R\Delta_{S^\prime}{}^S&\lambda_S-{\cal T}_{R^\prime RS}{}^K\mu_K
{\cal T}_{S^\prime}{}^{RSL}\lambda_L\Bigr)d\$ =-{i\over2}\oint_{\$}\mu^A
\gamma_A{}^K\lambda^BR_{KBcd}dx^c\wedge dx^d-\cr
+\oint_{\$}\mu^K\Bigl(&\delta^A_K\delta^B_R\Delta_{R^\prime}{}^L\lambda_L+
\varepsilon_{KR}{\cal T}_{R^\prime}{}^{ABL}\lambda_L\Bigr)\varepsilon^{
A^\prime B^\prime}T^{RR^\prime}{}_{AA^\prime BB^\prime}d\$.\cr}\eqno(7.4)
$$

\noindent The left hand side of these identities has surprisingly
symmetrical structure. If ${\mathop\pi_1}{}_{A^\prime}:=-i\Delta_{A^\prime}
{}^A$ $\mu_A$, ${\mathop\pi_2}{}_{B^\prime}:=-i\Delta_{B^\prime}{}^B
\lambda_B$ and their spinor components are defined by $\pi_{A^\prime}
=:\bar o_{A^\prime}\pi_{1^\prime}-\bar\iota_{A^\prime}\pi_{0^\prime}$
then

$$
\eqalignno{\bar\gamma^{A^\prime B^\prime}\Delta_{A^\prime}{}^A\mu_A\Delta
_{B^\prime}{}^B\lambda_B&=-\bigl({\mathop\pi_1}{}_{0^\prime}{\mathop\pi_2}
{}_{1^\prime}+{\mathop\pi_1}{}_{1^\prime}{\mathop\pi_2}{}_{0^\prime}\bigr)
&(7.5)\cr
\varepsilon^{A^\prime B^\prime}\Delta_{A^\prime}{}^A\mu_A\Delta_{B^\prime}
{}^B\lambda_B&=-\bigl({\mathop\pi_1}{}_{0^\prime}{\mathop\pi_2}{}_{1^\prime}
-{\mathop\pi_1}{}_{1^\prime}{\mathop\pi_2}{}_{0^\prime}\bigr).&(7.6)\cr}
$$

\noindent The right hand side of (7.5) is the well known expression related
to Penrose's kinematical twistor (eq. 9.9.29 of [1]), while (7.6) to the
infinity twistor (eq. 9.9.27 of [1]). In fact, if $\lambda_B$ is a
solution of the 2-surface twistor equation then identity (7.3) reduces to
Tod's expression of the kinematical twistor [1,18,20]; while (7.4) is an
(as far as we know) new identity.\par \vfill \eject

{\lbf Acknowledgements}\par
\vskip 0.5truecm
\noindent I am greatful to Professor H. Urbantke and Dr. J. Frauendiener
for discussions and to Professor B.G. Schmidt for encouragement to complete
this work. This work was partially supported by the Hungarian Scientific
Reseach Fund grant OTKA 1815.\par
\vskip 1truecm

{\lbf References}\par
\vskip 0.5truecm

\item{[1]}  R. Penrose, W. Rindler, Spinors and Spacetime, Vol.2, Cambridge
            Univ. Press, 1986
\item{[2]}  R. Geroch, Asymptotic structure of spacetime, in {\it Asymptotic
            structure of spacetime}, Eds.: F.P. Esposito and L. Witten,
	     (Plenum Press, New York) 1977; R. Geroch, G.T. Horowitz,
	     Phys.Rev.Lett. {\bf 40} 203 (1978)
\item{[3]}  R. Beig, B.G. Schmidt, Commun.Math.Phys. {\bf 87} 65 (1982);
            A. Ashtekar, J.D. Romano, Class.Quantum Grav. {\bf 9} 1069 (1992)
\item{[4]}  P. Schoen, S.T. Yau, Commun.Math.Phys. {\bf 65} 45 (1979)
\item{[5]}  E. Witten, Commun.Math.Phys. {\bf 80} 381 (1981)
\item{[6]}  J.M. Nester, Phys.Lett. {\bf 83 A} 241 (1981); W. Israel,
            J.M. Nester, Phys.Lett. {\bf 85A} 259 (1981)
\item{[7]}  M. Ludvigsen, J.A.G. Vickers, J.Phys.A: Math.Gen. {\bf 14} L389
            (1981); M. Ludvigsen, J.A.G. Vickers, J.Phys.A: Math.Gen.
	     {\bf 15} L67 (1982)
\item{[8]}  A. Ashtekar, G.T. Horowitz, Phys.Lett. {\bf 89A} 181 (1982);
            G.T. Horowitz, The positive energy theorem and its extensions,
	     in {\it Asymptotic Behavior of Mass and Spacetime Geometry},
	     {\it Lecture Notes in Phys. no 202}, Ed F.J. Flaherty (New York:
	     Springer), 1984
\item{[9]}  A. Sen, J.Math.Phys. {\bf 22} 1781 (1981)
\item{[10]} O. Reula, J.Math.Phys. {\bf 23} 810 (1982); O. Reula, K.P.
            Tod, J.Math.Phys. {\bf 25} 1004 (1984)
\item{[11]} A. Komar, Phys.Rev. {\bf 113} 934 (1959); J. Winicour, L.
            Tamburino, Phys.Rev.Lett. {\bf 15} 601 (1965); J.N. Goldberg,
	     in {\it General Relativity and Gravitation}, Ed.: A. Held,
	     Plenum Press, New York 1980; J. Katz, Class.Quantum Grav.
	     {\bf 2} 423 (1985)
\item{[12]} S.W. Hawking, J.Math.Phys. {\bf 9} 598 (1968); G.T. Horowitz,
            B.G. Schmidt, Proc.Roy.Soc. Lond. {\bf A 381} 215 (1982);
	     D. Christodoulou, S.-T. Yau, Some remarks on the quasi-local
            mass, in {\it Mathematics and General Relativity} (Contemporary
            Mathematics {\bf 71}) Ed. J.A. Isenberg (New York: AMS)
\item{[13]} S.A. Hayward, Quasi-local gravitational energy, preprint 1992,
            Max Planck Institute f\"ur Astrophysik
\item{[14]} R. Bartnik, Phys.Rev.Lett. {\bf 62} 2346 (1989)
\item{[15]} J.D. Brown, J.W. York,Jr, Phys.Rev.D. {\bf 47} 1407 (1993);
            S. Lau, Class.Quantum Grav. {\bf 10} 2379 (1993)
\item{[16]} J. Katz, D. Lynden-Bell, W. Israel, Class.Quantum Grav. {\bf 5}
            971 (1988)
\item{[17]} R. Kulkarni, V. Chellathurai, N. Dadhich, Class.Quantum Grav.
            {\bf 5} 1443 (1988)
\item{[18]} R. Penrose, Proc.Roy.Soc.Lond. {\bf A 381} 53 (1982);
            W.T. Shaw, Proc.Roy.Soc.Lond. {\bf A 390} 191 (1983);
	     K.P. Tod, Proc.Roy.Soc.Lond. {\bf A 388} 457 (1983);
	     B.P. Jeffryes, Class.Quantum Grav. {\bf 3} 841 (1986);
	     W.T. Shaw, Class.Quantum Grav. {3} 1069 (1986);
	     R.M. Kelly, K.P. Tod, N.M.J. Woodhouse, Class.Quantum Grav.
	     {\bf 3} 1151 (1986); N.M.J. Woodhouse, Class.Qu\-ant\-um Grav.
	     {\bf 4} L121 (1987); L.J. Mason, Class.Quantum Grav. {\bf 6} L7
	     (1989); A.D. Helfer, Class.Qu\-ant\-um Grav. {\bf 9} 1001 (1992)
\item{[19]} J.N. Goldberg, Conserved quantities at spatial and null infinity:
            The Penrose potential, Syracuse preprint (1989)
\item{[20]} K.P. Tod, Penrose's Quasi-local Mass, in {\it Twistors in
            Mathematics and Physics}, (London Math. Soc. Lecture Notes No.
	     156) Ed.: T.N. Bailey and R.J. Baston, (Cambridge Univ. Press,
	     Cambridge) 1990
\item{[21]} L.J. Mason, J. Frauendiener, The Sparling 3-form, Ashtekar
            Variables and Quasi-local Mass, in {\it Twistors in Mathematics
            and Physics}, (London Math. Soc. Lecture Notes No. 156) Ed.:
	     T.N. Bailey and R.J. Baston, (Cambridge Univ. Press, Cambridge)
	     1990
\item{[22]} M. Ludvigsen, J.A.G. Vickers, J.Phys.A: Math.Gen. {\bf 16} 1155
            (1983); G. Bergqvist, M. Ludvigsen, Class.Quantum Grav. {\bf 4}
	     L29 (1987); S.T. Swift, Class.Quantum Grav. {\bf 9} 1829 (1992)
\item{[23]} G. Bergqvist, M. Ludvigsen, Class.Quantum Grav. {\bf 6} L133
            (1988); G. Bergqvist, M. Ludvigsen, Class.Quantum Grav. {\bf
	     8} 697 (1991); G. Harnett, Class.Quantum Grav, {\bf 10} 407 (1993)
\item{[24]} A.J. Dougan, L.J. Mason, Phys.Rev.Lett. {\bf 67} 2119 (1991);
            A.J. Dougan, Class.Quantum Grav. {\bf 9} 2461 (1992);
            L.B. Szabados, Class.Quantum Grav. {\bf 10} 1899 (1993)
\item{[25]} G. Bergqvist, Class.Quantum Grav. {\bf 9} 1917 (1992)
\item{[26]} G. Bergqvist, Class.Quantum Grav. {\bf 9} 1753 (1992)
\item{[27]} R. Geroch, A. Held, R. Penrose, J.Math.Phys. {\bf 14} 874 (1973)
\item{[28]} R. Penrose, W. Rindler, Spinors and Spacetime, Vol.1, Cambridge
            Univ. Press, 1984
\item{[29]} S.A. Hugget, K.P. Tod, An Introduction to Twistor Theory,
            (London Mathematical Society Texts 4), Cambridge University
	     Press, Cambridge 1985
\item{[30]} L.B. Szabados, Two Dimensional Sen Connections, in {\it Proc.
            4th Hungarian Relativity Workshop, July 12-17, 1992, G\'ardony},
            Eds.: R.P. Kerr and Z. Perj\'es, (to be published)
\item{[31]} S. Kobayashi, K. Nomizu, Foundations of Differential Geometry,
            Vol 1, Interscience, 1964
\item{[32]} S. Kobayashi, K. Nomizu, Foundations of Differential Geometry,
            Vol 2, Interscience, 1968
\item{[33]} J. Frauendiener, Class.Quantum Grav. {\bf 8} 1881 (1991)
\item{[34]} R.J. Baston, Twistor Newsletter, {\bf 17} 31 (1984)

\end